\documentclass[%
 aip,
 rsi,%
 amsmath,amssymb,
 preprint,%
]{revtex4-1}

\usepackage{graphicx}
\usepackage{dcolumn}
\usepackage{bm}
\usepackage[caption=false]{subfig} 
\usepackage{float}

\begin{document}


\title{Standing-wave-excited multiplanar fluorescence in a laser scanning microscope reveals 3D information on red blood cells}

\author{Rumelo Amor$^{*}$}
 \email{rumelo.c.amor@strath.ac.uk}
\affiliation{ 
Centre for Biophotonics, Strathclyde Institute of Pharmacy and Biomedical Sciences, University of Strathclyde, 161 Cathedral Street, Glasgow G4 0RE, United Kingdom
}%

\author{Sumeet Mahajan}
\affiliation{%
Institute of Life Sciences and Department of Chemistry, University of Southampton, Highfield Campus, Southampton SO17 1BJ, United Kingdom
}%

\author{William Bradshaw Amos}
\affiliation{%
MRC Laboratory of Molecular Biology, Francis Crick Avenue, Cambridge Biomedical Campus, Cambridge CB2 2QH, United Kingdom
}%

\author{Gail McConnell}
\affiliation{ 
Centre for Biophotonics, Strathclyde Institute of Pharmacy and Biomedical Sciences, University of Strathclyde, 161 Cathedral Street, Glasgow G4 0RE, United Kingdom
}%

\date{\today}

\begin{abstract}
Standing-wave excitation of fluorescence is highly desirable in optical microscopy because it improves the axial resolution. We demonstrate here that multiplanar excitation of fluorescence by a standing wave can be produced in a single-spot laser scanning microscope by placing a plane reflector close to the specimen. We report here a variation in the intensity of fluorescence of successive planes related to the Stokes shift of the dye. We show by the use of dyes specific for the cell membrane how standing-wave excitation can be exploited to generate precise contour maps of the surface membrane of red blood cells, with an axial resolution of $\approx$90 nm. The method, which requires only the addition of a plane mirror to an existing confocal laser scanning microscope, may well prove useful in studying diseases which involve the red cell membrane, such as malaria.
\end{abstract}

\maketitle

Standing-wave excitation of fluorescence has been achieved previously using two methods: the 4Pi geometry, where counter-propagating coherent waves are directed into the specimen \cite{Hell92a, Hell92b, Gu94}, and the simpler approach of placing a reflector close to the specimen, done in wide-field illumination \cite{Lanni86, Lanni86p, Bailey93}, and used recently in laser-scanning \cite{Elsayad2013}. Here, we have observed that the standing waves close to a reflector are radially modulated in a precise way, and show that this corresponds to a moir\'{e} pattern between the excitation and emission standing-wave fields. We demonstrate that standing-wave-excited fluorescence in red blood cells allows precise contour-mapping of the cell membrane, with an axial resolution of $\approx$90 nm, compared with the 300 nm of a confocal microscope \cite{Amos2012}. 

In standing-wave fluorescence microscopy using a reflector, interference can occur between the direct and reflected components of the excitation beam. This creates a standing wave pattern that excites fluorescence only at the antinodal planes, which are regularly spaced in the axial direction. Using an excitation wavelength $\lambda$ and a medium of refractive index $n$, and given that the antinodal planes are parallel to the reflector surface, the axial resolution is improved to the sine squared intensity peaks at the antinodes, with a full width at half maximum of $\lambda/4n$ \cite{Lanni86, Lanni86p, Bailey93}. However, the drawback in using standing-wave excitation is that if the specimen is thicker than the antinodal spacing $\lambda/2n$, several planes may be excited simultaneously within the depth of field of the objective lens and their contributions may be hard to separate from each other \cite{Egner2005, Freimann97}. But if the specimen is simple in structure as in a cell membrane and the height variation is not too rapid across the field of the microscope, distinct excitation fringes can be recognized in xy images and used as contour lines for precise three-dimensional mapping.

\section*{Results}
\begin{figure}[H]
\centering
\includegraphics[width=0.85\textwidth]{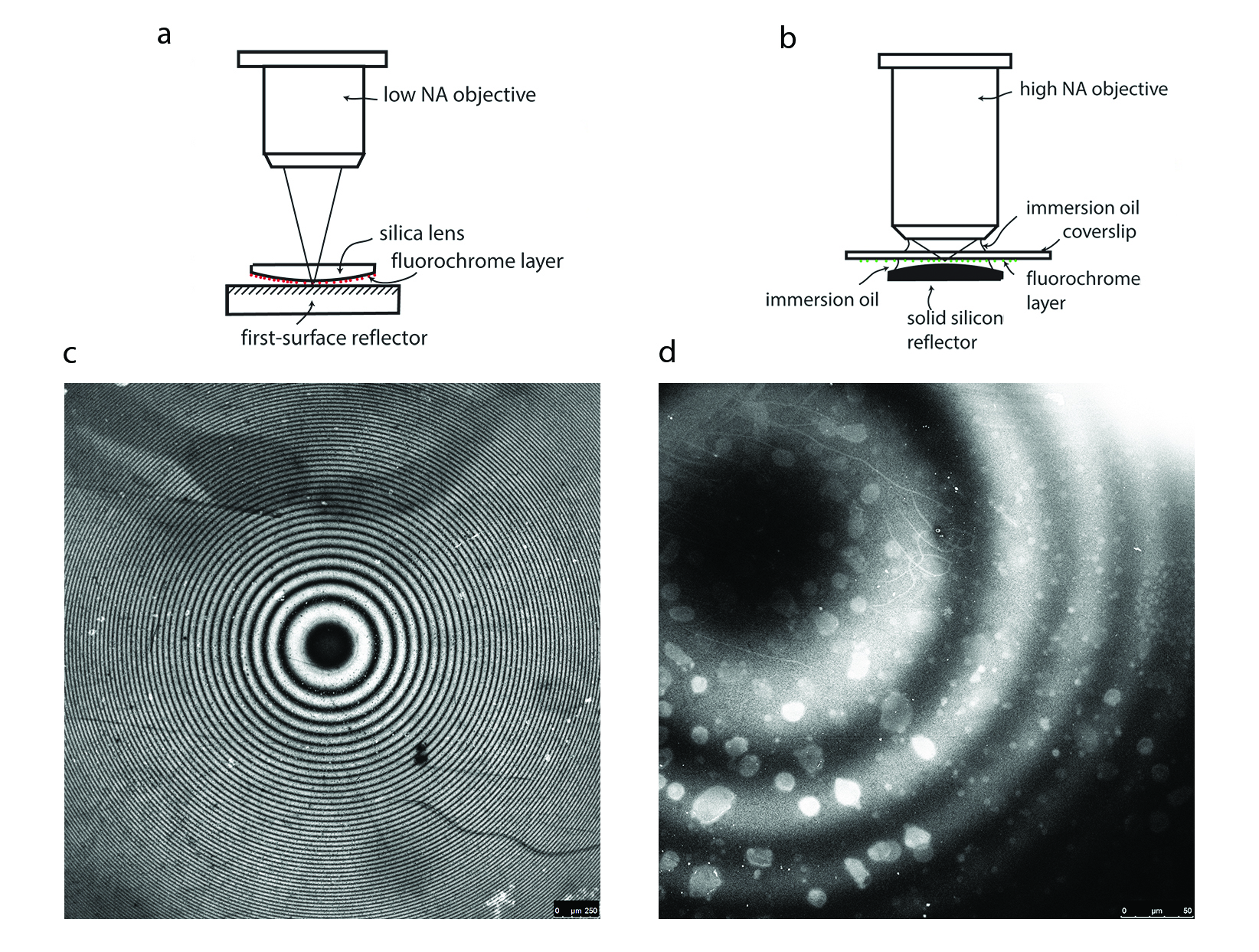}

\centering
\captionsetup{justification=raggedright,
singlelinecheck=false
}
\caption{Experimental setup and fluorescence images obtained with varying distance between fluorochrome and reflector. Setup (\emph{a}) was used to investigate fluorescence in air with dry objectives of low numerical aperture and long working distance (e.g. 5x N.A. 0.15). The fluorochrome, e.g. Atto 532 dye, was deposited as a thin layer on the convex side of a planoconvex lens, which was balanced on an aluminised first-surface reflector. In certain experiments the broad-band reflector was replaced with a plate coated to reflect only the excitation wavelengths or only the emission wavelengths of the dye. For studies at high numerical aperture, setup (\emph{b}) was used, allowing the fluorochrome to be brought within the limited working distance of the high-N.A. lens and for normal immersion oil and a correctly specified coverslip to be used. The dye in this case was the sodium salt of fluorescein deposited in a thin layer on the underside of the coverslip and the convex surface of a solid silicon planoconvex lens was used as the reflector, since it behaves as such in visible wavelengths. (\emph{c}) Fluorescence image from the setup in (\emph{a}), showing concentric fringes from multiple planes of excitation at the standing-wave antinodes. Scale bar=250 $\mu$m. (\emph{d}) Fluorescence fringes from the setup in (\emph{b}), consistent with standing-wave excitation in spite of the presence of rays at a wide range of angles of incidence. Scale bar=50 $\mu$m.}\label{Fig1} 
\end{figure}

\begin{figure}[H]
\centering
\includegraphics[width=\textwidth]{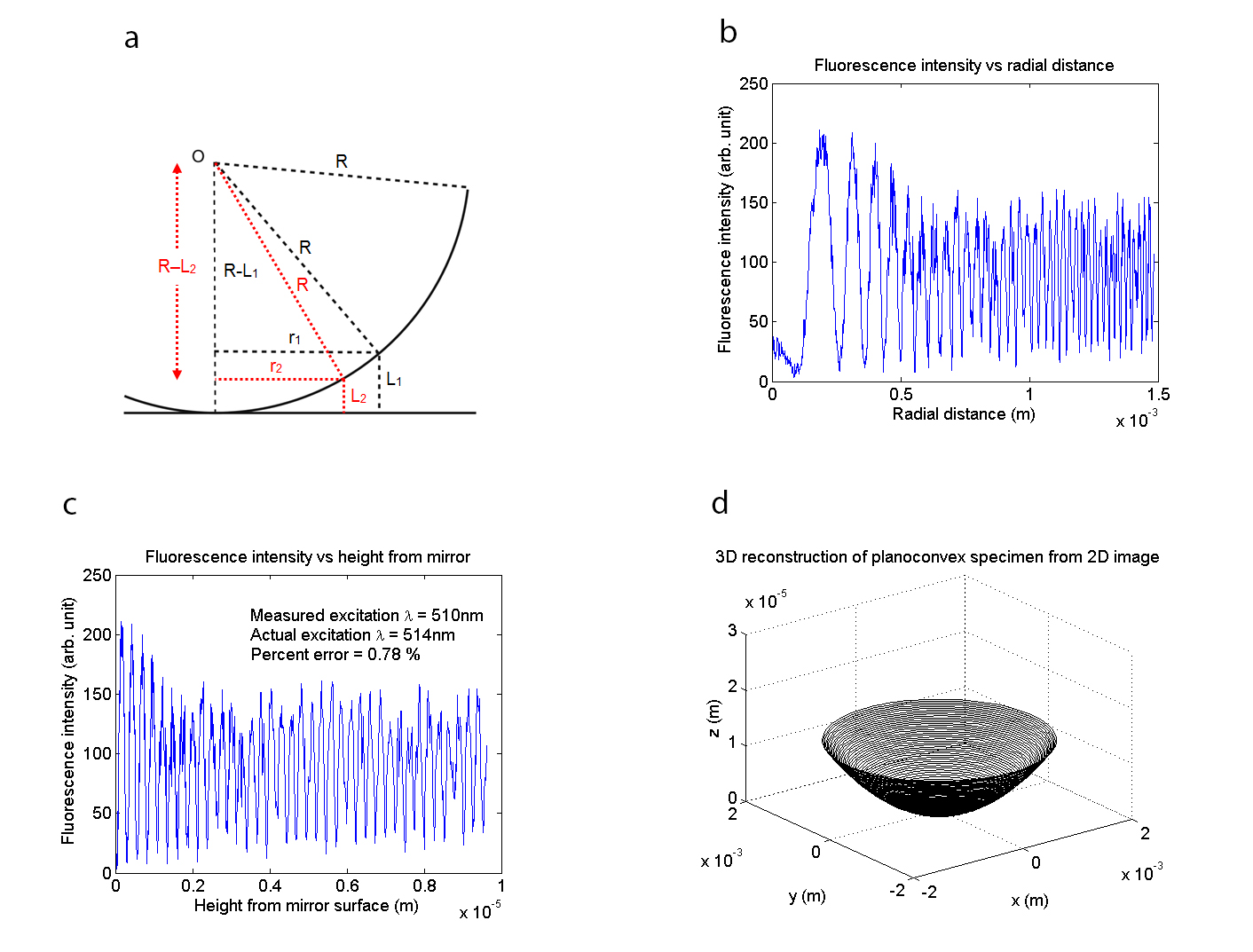}

\centering
\captionsetup{justification=raggedright,
singlelinecheck=false
}

\caption{Three-dimensional reconstruction from a single two-dimensional image. (\emph{a}) Diagram of planoconvex specimen in contact with the mirror. The height $L_1-L_2$ separating two successive bright fringes of radii $r_{1}$ and $r_{2}$ is given by $\sqrt{R^{2}-r_{2}^{2}}-\sqrt{R^{2}-r_{1}^{2}}$. (\emph{b}) Fluorescence intensity in Fig.~\ref{Fig1}c plotted versus radial distance, showing the fringe spacing getting smaller with distance from the centre, consistent with a fluorescent shell of spherical shape excited by the evenly-spaced antinodes of a standing wave. (\emph{c}) Fluorescence intensity plotted versus height from the mirror surface, obtained using the geometry in (\emph{a}), showing evenly-spaced peaks where the dye cut the antinodes of the standing wave. The measured antinodal spacing is 255 nm, accurate to within 0.78 $\%$ of the actual value of $\lambda$/2n=257 nm using an excitation wavelength of 514 nm and a refractive index of n=1 (air). (\emph{d}) Three-dimensional reconstruction of planoconvex specimen from the two-dimensional fluorescence image in Fig.~\ref{Fig1}c.}\label{Fig2} 
\end{figure}

\noindent
\textbf{Model specimens.} In order to investigate the intensity envelope of multiple planes of emission, we first constructed a model specimen consisting of a monolayer of Atto 532 dye covalently attached to the surface of a planoconvex silica lens which we placed on top of a broad-band dielectric mirror, as in Fig.~\ref{Fig1}a. We performed single-spot laser-scanning on this specimen with a Leica SP5 confocal microscope using a 5x/0.15 N.A. dry objective and observed concentric fluorescence fringes as in Fig.~\ref{Fig1}c, with the fringe spacing as one would expect if the fluorescent monolayer cut through successive antinodal planes of standing-wave excitation. The centre of the fringes, which is the point of contact between the specimen and the mirror, is dark as expected because the mirror surface is the first node of the standing wave. This result is the fluorescence equivalent of classic demonstrations of standing light waves \cite{Wiener1890, Ditchburn58, Jenkins76, Hecht98}. We sought to extend this experiment to include objectives of high numerical aperture, such as are needed in cell biology imaging. These have a limited working distance of a fraction of a millimetre. We coated a glass coverslip with a thin layer of fluorescein dye and placed this on top of a planoconvex silicon lens (\#69-673, Edmund Optics, Inc., USA) which served as a mirror, as in Fig.~\ref{Fig1}b. We performed laser-scanning in this setup using a 40x/1.30 N.A. oil immersion objective and again obtained a pattern of concentric fluorescence fringes, shown in Fig.~\ref{Fig1}d, consistent with standing-wave excitation, in spite of the presence of rays at a wide range of angles of incidence.  We discuss the origin of apparently planar antinodal zones in a single-point scanning microscope even with objectives of high numerical aperture (where the focussed spot produces curved antinodes) in Supplementary Section S2 and Supplementary Fig. S2. To verify that the axial fringe spacing is constant and comparable to the antinode spacing $\lambda/2n$, we have plotted the fringe intensity as a function of axial height of the fluorochrome above the reflector surface. Figure~\ref{Fig2}a shows a median section of the half-spherical surface of the specimen in Fig.~\ref{Fig1}a, of radius $R$, centred at $O$, in contact with the mirror. A bright fringe of radius $r_1$ is at a height $L_1$ above the mirror, and the fringe next it, of radius $r_2$, is at a height $L_2$. The height separating these two successive fringes is then obtained from the two triangles formed separately by $r_1$ and $r_2$, the lengths of which are given by the Pythagorean theorem as $R^2=r_{1}^{2}+(R-L_1)^2$ and $R^2=r_{2}^{2}+(R-L_2)^2$. The separation height is therefore $L_1-L_2=\sqrt{R^2-r_{2}^{2}}-\sqrt{R^2-r_{1}^{2}}$. Taking an intensity profile along a radial line in Fig.~\ref{Fig1}c, we are able to plot the fluorescence intensity as a function of radial distance, as in Fig.~\ref{Fig2}b. A bright fringe in Fig.~\ref{Fig1}c corresponds to a peak in Fig.~\ref{Fig2}b, which shows the fringe spacing getting smaller with distance from the centre, which is what we would expect if a fluorescent shell of spherical shape is excited by the evenly-spaced antinodes of a standing wave. Using our result above for the height $L_1-L_2$ separating two successive fringes, expressed in terms of the radial distance, we are then able to translate the plot in Fig.~\ref{Fig2}b into one of fluorescence intensity as a function of height from the mirror surface, as in Fig.~\ref{Fig2}c, where we have obtained evenly-spaced peaks of fluorescence emission, corresponding to where the dye is excited by successive antinodes of the standing wave. In the plot there are 37 peaks, and by locating the positions of the maxima by flipping the function vertically and using \emph{fminbnd} in MATLAB$^\circledR$, which finds the minimum of a function within an interval, we find that the antinodal spacing is 255 nm. In the experiment setup, there is only air between the specimen and the mirror ($n=1$), and using $\lambda=514$ nm for the excitation, we obtain the actual antinode spacing $\lambda/2n=257$ nm, showing that our experimental result is within 0.78 $\%$ of the actual value. Finally, the measured experimental antinodal spacing in Fig.~\ref{Fig2}c and the actual lateral separation of the fringes in Fig.~\ref{Fig2}b lead to an accurate reconstruction of the planoconvex specimen in three dimensions, shown in Fig.~\ref{Fig2}d. 

\begin{figure}[H]
\centering
\includegraphics[width=0.95\textwidth]{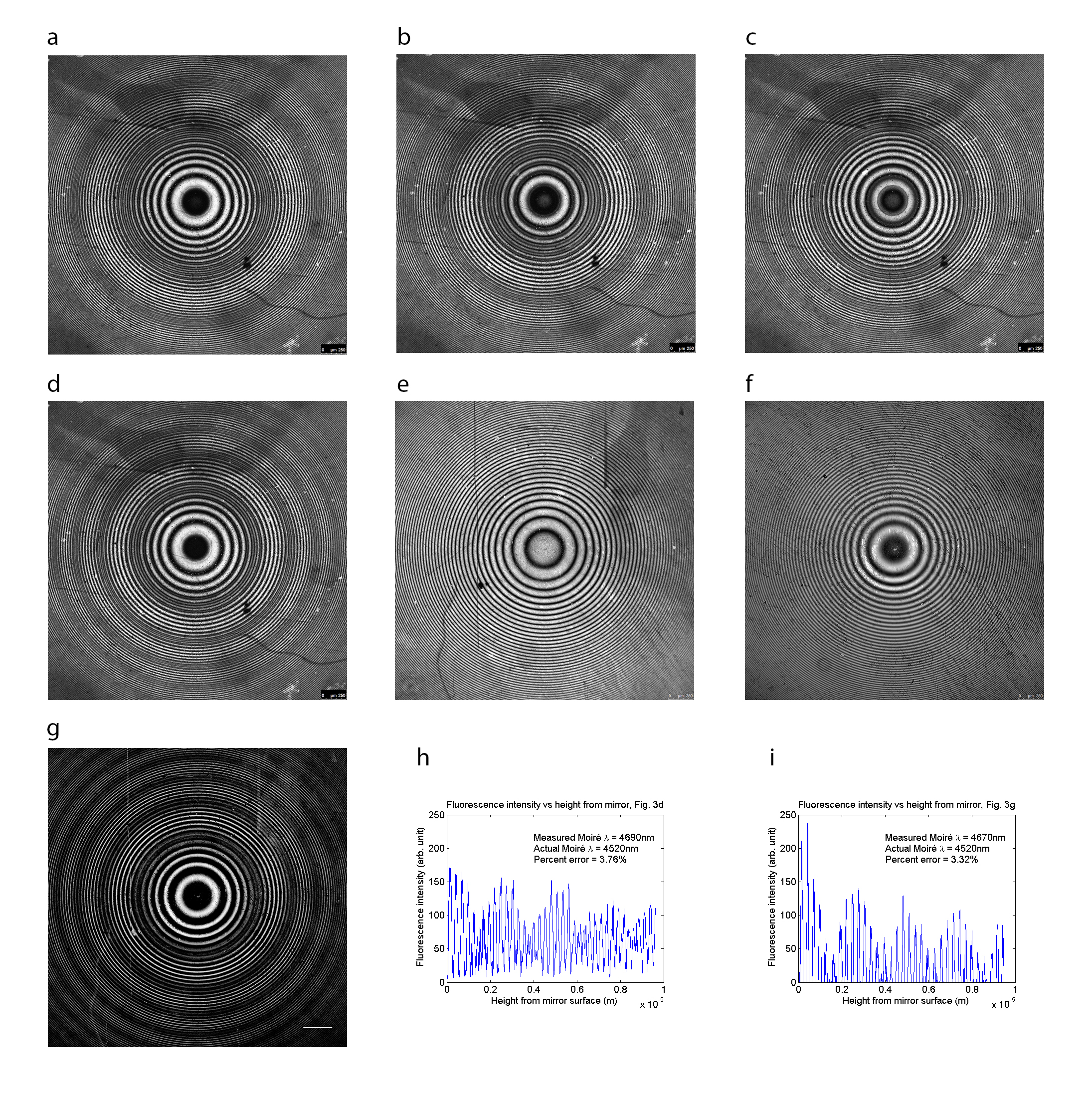}

\centering
\captionsetup{justification=raggedright,
singlelinecheck=false
}

\caption{Moir\'{e}-like radial modulation of fluorescence fringe brightness observed at different detection wavelengths and also simulated. Detecting the fluorescence emission at a bandwidth of 5 nm centred at (\emph{a}) 550 nm, (\emph{b}) 560 nm, (\emph{c}) 570 nm, and (\emph{d}) 580 nm shows the directly-observed fringe pattern to be radially modulated, with a frequency proportional to the Stokes shift, consistent with a moir\'{e} pattern between the excitation and emission standing-wave fields. (\emph{e}) In the absence of the excitation standing-wave field, the modulation is absent in the 580 nm emission. (\emph{f}) Excitation standing-wave field fringe pattern. (\emph{g}) Subtraction of the emission-only fringe pattern (\emph{e}) from the excitation fringe pattern (\emph{f}) gives a modulated image identical to the moir\'{e} pattern in (\emph{d}). The measured spatial periods in (\emph{d}) and (\emph{g}) are: (\emph{h}) 4690 nm, and (\emph{i}) 4670 nm, respectively, both within 4 $\%$ of the theoretical value of 4520 nm for a moir\'{e} pattern between 514 nm and 580 nm. Scale bar=250 $\mu$m.}\label{Fig3_beats} 
\end{figure}

In addition to fluorescence fringes detected at a wide spectral bandwidth of 100 nm, we also observed that the fringes were radially modulated in a precise way particularly when the detection bandwidth was reduced to 5 nm, which is the spectral resolution of the Leica SP$^\circledR$ spectral detector \cite{Leica5nm2014}. In our experiments, the modulation has a frequency proportional to the Stokes shift, consistent with a moir\'{e} pattern between the excitation and emission, which occurs at the difference frequency between the periodicity of the excitation standing-wave field and the periodicity of the standing-wave field created by the light at the selected emission wavelength. A moir\'{e} pattern arising from the presence of two distinct patterns has a spatial frequency that is the difference frequency between the two individual patterns \cite{Rogers77, Myers2004, Stupp2008}. Figure~\ref{Fig3_beats} shows the moir\'{e} pattern in the directly-observed fringes and clearly demonstrates the increasing frequency as the fluorescence is detected from 550 nm (Fig.~\ref{Fig3_beats}a) through to 580 nm (Fig.~\ref{Fig3_beats}d). Taking Fig.~\ref{Fig3_beats}d as an example, a moir\'{e} pattern from 514 nm excitation and 580 nm emission has a theoretical spatial period of $L=4520$ nm from $1/L=~$1/514 nm -- 1/580 nm. Fig.~\ref{Fig3_beats}h plots the fluorescence intensity in Fig.~\ref{Fig3_beats}d as a function of height from the mirror surface, using the method described for Fig.~\ref{Fig2}. Using the spacing between peaks in the moir\'{e} pattern, equal to $L/2n$, we measured a spatial period of 4690 nm, accurate to within 3.8 $\%$ of the theoretical value. Previous work has shown that when fluorescent molecules are placed close to a mirror, the emitting molecule acts as an oscillating dipole (an antenna) and that the reflected and unreflected parts of the emitted fluorescence wave interfere with each other, producing fringes from wide-angle interference \cite{Selenyi11, Selenyi38, Selenyi39} and oscillations in the fluorescence decay time \cite{Drexhage68, Drexhage70a, Drexhage70b, Drexhage74}, with analytic expressions for the radiation patterns derived from electrodynamic theory \cite{Lukosz77i, Lukosz77ii, Lukosz79} and modified further from the fixed-dipole amplitude assumption to model the fluorophore as a dipole of constant power and variable amplitude \cite{Axelrod87, Axelrod89}. Subsequent work in this area studied the modulation of fluorescence intensity with distance from the mirror, and explained the modulation as a consequence of the interference effects of the excitation and of the emission being both present \cite{Lambacher96}. We have obtained the fluorescence fringes from the standing-wave field created by the self-interference of the emission alone, by replacing the mirror with a laser-line notch filter (LL01-514-25, Semrock, Inc., USA) to transmit the excitation and therefore suppress the excitation standing-wave field. As shown in Fig.~\ref{Fig3_beats}e, the modulation in the fringes detected at 580 nm is absent in this case. More importantly, as can be seen in Fig.~\ref{Fig3_beats}g, subtraction of the 580 nm emission fringe pattern in Fig.~\ref{Fig3_beats}e from the fringe pattern of the excitation in Fig.~\ref{Fig3_beats}f gives a modulated fringe pattern identical to the moir\'{e} pattern in Fig.~\ref{Fig3_beats}d. The spatial period of this modulation, as measured using the peak spacing in Fig.~\ref{Fig3_beats}i, is 4670 nm, which is within 3.3 \% of the theoretical value, confirming that it occurs at the difference frequency between the excitation and emission. The observation of a moir\'{e} pattern of the type we describe here appears to be novel in optical fluorescence microscopy, and suggests a method by which the successive standing wave antinodes could be distinguished by their amplitude. It may be possible to extend this distinction further to an unequivocal identification by analysis of fringe colour, using multiple colocalised fluorochromes with overlapping excitation spectra but differing in emission wavelength and Stokes shift.

In order to test whether the fluorescence fringes that we observed were caused by standing-wave excitation formed by the interaction of the excitation light with the mirror, and not by a Fabry-Perot cavity effect, where the silica-air interface on one side and the mirror on the other formed the cavity, we constructed specimens in which we put immersion oil ($n=1.52$), pure glycerol ($n=1.47$) and a solution of 91 $\%$ glycerol in water ($n=1.46$) \cite{Dow2014} between the substrate and the mirror in Fig.~\ref{Fig1}a. Putting a film of fluid that has the same refractive index as the silica lens ($n=1.46$) \cite{Comar2014} eliminates reflection from the convex surface. We still observed the concentric fluorescence fringes in all these cases, shown in Supplementary Fig. S3, proving that the fringe pattern is caused by a standing wave only.\\

\noindent
\textbf{Red blood cells.} Recent work has shown elegantly that the first antinode can be positioned near the contact between living cells and their substrate, by growing cells on a dielectric layer of appropriate optical thickness deposited on a reflector \cite{Elsayad2013}. Since we wished to observe multiple antinodes, we selected as specimen red blood cells stained with the plasma membrane-specific dyes DiI \cite{Schlessinger77, Chang89, Chen2008, Hod2010, Agmon95}, DiO \cite{Chen2008, Hod2010} or DiI(5) \cite{Agmon95, Webb83}. We stabilised mouse red blood cells in their characteristic shape of a biconcave disk about 7-8 $\mu$m in diameter and about 1 $\mu$m thick at the thinnest point at the centre and 2.5 $\mu$m at the thickest point in the periphery \cite{Greer2009}. For this experiment, we obtained fresh blood by cardiac puncture from healthy mice. In order to exclude the possibility of lensing or focussing effects from refractile material inside the cell, we also prepared red blood cell ghost specimens in addition to intact red blood cell specimens. Red blood cell ghosts have the same shape as normal red blood cells but with the haemoglobin removed from the interior. We used the protocol of Harris \emph{et al.} \cite{Harris2001} for the preparation of the ghosts, which we describe in the \emph{Methods} section. We stained the membrane of the prepared ghosts and intact red blood cells with DiI, DiO or DiI(5), and mounted these on top of a mirror. We also prepared specimens of the ghosts and red blood cells on ordinary non-reflective glass microscope slides, to serve as controls. All specimens were mounted using 4 $\%$ bovine serum albumin (BSA) in phosphate-buffered saline solution ($n=1.34$) \cite{Barer54} to match the average refractive index of the membrane, and imaged with a Leica SP5 confocal microscope using a 100x/1.40 N.A. oil immersion objective. The excitation wavelength used was 543 nm for DiI, 488 nm for DiO and 633 nm for DiI(5), and the fluorescence signal was detected at 555--655 nm for DiI, 498--598 nm for DiO and 650--750 nm for DiI(5). 

\begin{figure}[H]
\centering
\includegraphics[width=\textwidth]{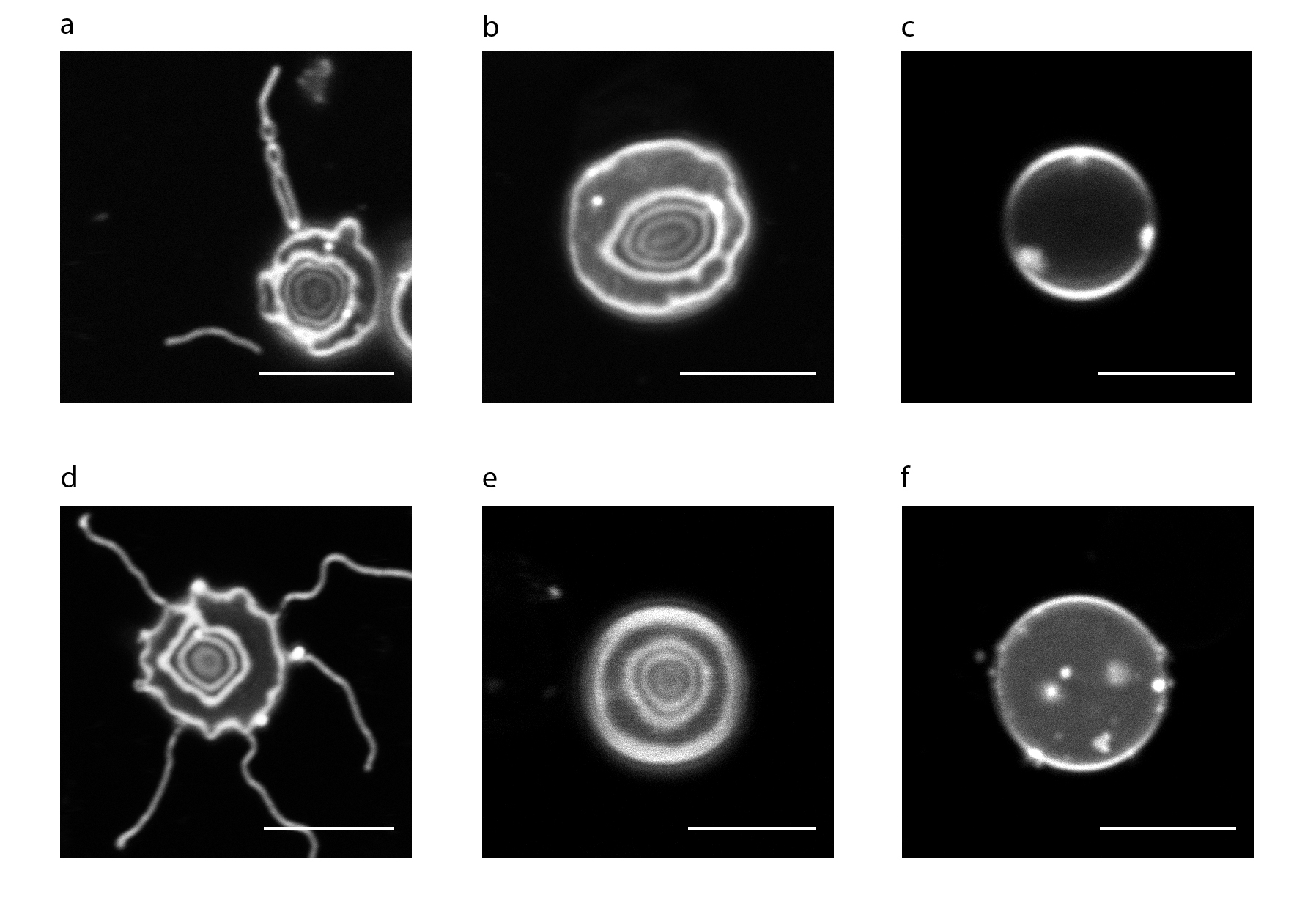}

\centering
\captionsetup{justification=raggedright,
singlelinecheck=false
}

\caption{Precise contour-mapping of the red blood cell membrane. Fringes in red blood cell ghosts (\emph{a}) and intact red blood cells (\emph{b}, \emph{d}, \emph{e}) from multiplanar excitation of fluorescence at the  standing-wave antinodes. The cells were stained with the membrane-specific dyes DiI (\emph{a} and \emph{b}), DiO (\emph{d}) or DiI(5) (\emph{e}), and mounted on top of a mirror using 4 $\%$ BSA in PBS (n=1.34) to match the average refractive index of the membrane. Using an excitation wavelength of 488 nm for DiO, the axial resolution from the full width at half maximum of the high-intensity excitation light at the antinodes is $\lambda/4n\approx\,$90 nm. Control specimens of red blood cell ghosts (\emph{c}) and intact red blood cells (\emph{f}) mounted on ordinary non-reflective glass microscope slides emit fluorescence only at the outline of the cell. Scale bar=5 $\mu$m.}\label{Fig4}
\end{figure}

With the ghosts on a slide, and without standing-wave excitation, we would expect to see, and did see, only a bright outline of the cell, as shown in Fig.~\ref{Fig4}c. With standing-wave excitation using a mirror, we saw a well-defined pattern of fringes, shown in Fig.~\ref{Fig4}a, as a result of multiple planes of fluorescence excitation due to the standing wave antinodes. With the specimens of intact red blood cells mounted on top of a mirror, we also observed multiple planes of fluorescence in the membrane, as shown in Fig.~\ref{Fig4}b for DiI, Fig.~\ref{Fig4}d for DiO and Fig.~\ref{Fig4}e for DiI(5). In the control without the mirror, the fluorescence again comes only from the cell outline, as shown in Fig.~\ref{Fig4}f. We show a differential interference contrast image of the red blood cells in Supplementary Fig. S4, which confirms that the cells are normal discocytes. Using 488 nm excitation, we have an axial resolution of $\lambda/4n\approx\,$90 nm from the full width at half maximum of the high-intensity excitation light at the antinodes. This ability to simultaneously visualise red blood cells at multiple planes reveals three-dimensional structures such as the long membrane protrusions visible in Figs.~\ref{Fig4}a~and~\ref{Fig4}d. We aim to test the value of this method for the study of acanthocytes and other abnormally-shaped red blood cells, such as occur in blood disorders \cite{Tse99, LimHW2002, Gallagher2005} and to study the mechanisms of parasite invasion and egress in diseases which involve the red cell membrane, such as malaria \cite{Dvorak75, Pinder98, Salmon2001, Stubbs2005, Gilson2009, Riglar2011}.

\section*{Methods}
\noindent
\textbf{Preparation of model specimens.} To covalently attach a monolayer of dye to planoconvex silica lenses (63 PS 16 and 250 PS 25, Comar Instruments, UK), the lenses were rinsed thoroughly with dry acetone and dipped in a 2 $\%$ mass concentration solution of 3-amino-propyltrimethoxysilane (APTMS, Sigma-Aldrich, USA) in dry acetone. The optimum time for coating the silica surface with APTMS was found to be 6 hours. This is important to ensure near-monolayer attachment of the dye to the silica surface as APTMS can polymerise on the surface forming multiple layers, especially in the presence of water. After this time the lenses were taken out, rinsed with dry acetone a few times and gently blow-dried with nitrogen. A 10 $\mu$M solution of Atto 532 NHS ester (88793, Sigma-Aldrich, USA) was prepared in pH 8.1, 0.01 M phosphate buffer. The lenses were soaked in this solution in a vial, sealed with parafilm and left overnight in a dark chamber wrapped in aluminium foil. Thereafter, the remaining solution was recovered and the silica surfaces thoroughly rinsed with deionised water. The lenses were then blow-dried with nitrogen and stored in the dark to prevent photodamage. 

To coat glass coverslips with a thin film of fluorescein dye, fluorescein was dissolved in ethanol at a dilute concentration of 10 $\mu$M. A 5 $\mu$L drop of this solution was placed on the coverslips and allowed to evaporate, producing a thin coating of fluorescein on the glass.\\

\noindent
\textbf{Preparation of red blood cell ghost specimens and intact red blood cell specimens.} Fresh mouse blood was collected by cardiac puncture, yielding about 0.5 mL blood from one mouse. The blood was immediately mixed with the anti-coagulant acid citrate dextrose (ACD) to prevent clotting, at a ratio of 1 part ACD to 3 parts blood. 100 mL of ACD is 1.32 g trisodium citrate (BP327-500, Fisher Scientific, USA), 0.48 g citric acid (423565000, Acros Organics, USA), 1.40 g dextrose (BP350-500, Fisher Scientific, USA) and distilled water to make up 100 mL of solution. The red blood cells were washed three times in phosphate-buffered saline solution (PBS: 137 mM NaCl, 2.7 mM KCl, 10.6 mM Na$_{2}$HPO$_{4}$, 8.5 mM KH$_{2}$PO$_{4}$, pH 7.4) and suspended to their original hematocrit in modified balanced salt solution (MBSS: 134 mM NaCl, 6.2 mM KCl, 1.6 mM CaCl$_{2}$, 1.2 mM MgCl$_{2}$, 18.0 mM HEPES, 13.6 mM glucose, pH 7.4, 37 $^\circ$C). 200 $\mu$L of this suspension was set aside, resuspended in 4 $\%$ BSA (A7906, Sigma-Aldrich, USA) in PBS, and stored at 4 $^\circ$C overnight for the preparation of intact red blood cell specimens. The washed red blood cells were suspended in 10 mL of PBS diluted 2/5 in distilled water with 1 mM CaCl${_2}$. The lower osmotic pressure causes the cells to rupture and release their haemoglobin, creating ghosts. After 30 minutes at 0 $^\circ$C, 5-fold concentrated PBS containing 1 mM CaCl$_{2}$ was added to the ghost preparation to restore isotonicity. The suspension was then incubated for 45 minutes at 37 $^\circ$C to reseal the membrane, and sealed ghosts were collected by centrifugation at 2,500$\times$g for 10 minutes. The ghosts were washed in PBS until the supernatant appeared free from haemoglobin and resuspended in 1 ml of 4 $\%$ BSA in PBS and stored at 4 $^\circ$C overnight. The ghosts and intact red blood cells were stained for the membrane by adding a 1 mg/ml stock solution of DiI or DiO (D-282 and D-275, Invitrogen, Ltd., UK) in ethanol, or a 1 mg/ml stock solution of DiI(5) (D-307, Invitrogen, Ltd., UK) in dimethyl sulfoxide (D8418, Sigma-Aldrich, USA), at a ratio of 1 part stock solution to 99 parts cell suspension, and the suspension was incubated for 30 minutes at 37 $^\circ$C. The ghosts and cells were then spun down and rinsed with PBS and resuspended in 4 $\%$ BSA in PBS. To promote the adhesion of ghosts and cells to the surface of a mirror or microscope slide, these surfaces were coated with poly-L-lysine. A working solution of poly-L-lysine was made by preparing a 1 in 10 dilution of 0.1 $\%$ mass concentration poly-L-lysine stock solution (P8920, Sigma-Aldrich, USA) in water. The mirrors and microscope slides were sterilised in 95 $\%$ ethanol and dried before coating. These were submerged in poly-L-lysine in a sterile Petri dish, incubated for 15 minutes at 37 $^\circ$C, washed three times with PBS, and placed under ultraviolet light in a cell culture hood for 15 minutes to sterilise. Specimens of red blood cell ghosts and intact red blood cells were then prepared by pipetting 5 $\mu$L of suspension onto a mirror or microscope slide and lowering a cover slip gently over the drop at an angle, allowing the liquid to spread out.\\

\noindent
\textbf{Confocal laser scanning microscope settings.} Laser scanned imaging of all specimens was performed on a Leica SP5 DM600 confocal microscope, using a 5x/0.15 N.A. HCX PL FLUOTAR DRY objective for Fig.~\ref{Fig1}c and Fig.~\ref{Fig3_beats}, a 40x/1.30 N.A. HCX PL APO CS OIL objective for Fig.~\ref{Fig1}d, and a 100x/1.40 N.A. HCX PL APO CS OIL objective for Fig.~\ref{Fig4}. A schematic diagram of the experimental setup is shown in Supplementary Fig. S1. For Fig.~\ref{Fig1}c and Fig.~\ref{Fig3_beats}, the excitation came from the 514 nm line of an Argon laser, with an average power of about 5 $\mu$W at the specimen for Fig.~\ref{Fig1}c, and 30--90 $\mu$W for Fig.~\ref{Fig3_beats}, the fluorescence in the latter case being collected at a narrow bandwidth of 5 nm to detect the beats pattern. The image size was 2048$\times$2048 pixels averaged over 8 lines of scan, which took about 40 seconds per frame. For Fig.~\ref{Fig1}d, the excitation was from the 488 nm line of the Argon laser, with an average power of 20 $\mu$W at the specimen. The image size was 2048$\times$2048 pixels averaged over 8 lines and took 40 seconds per frame. For the imaging of red blood cell ghosts and intact red blood cells in Fig.~\ref{Fig4}, the excitation was from the 543 nm line of a Helium-neon laser with an average power of about 1--3 $\mu$W at the specimen, except in Fig.~\ref{Fig4}d, where the excitation was from the 488 nm line from the Argon laser with an average power of about 3 $\mu$W at the specimen, and Fig.~\ref{Fig4}e, where the excitation was from the 633 nm line of a Helium-neon laser with an average power of about 5 $\mu$W at the specimen. All panels had a size of 1024$\times$1024 pixels averaged over 8 lines and took about 20 seconds per frame, cropped further to 480$\times$480 pixels.


\subsection*{Acknowledgements}

This work was supported by the Scottish Universities Physics Alliance (SUPA) and the Engineering and Physical Sciences Research Council (EPSRC) (grant no. EP/I006826/1). R.A. is supported by a SUPA INSPIRE studentship. We thank J. Crowe and O. Millington (University of Strathclyde) for the collection of mouse whole blood and helpful discussions on the preparation of red blood cell ghost specimens.


\subsection*{Author contributions}

W.B.A. and G.M. initiated and supervised the project. S.M. covalently attached Atto 532 dye to the planoconvex silica specimens. R.A. prepared the fluorescein, intact red blood cell and red blood cell ghost specimens, performed the experiments, and wrote the manuscript. All authors contributed to analysing the data and editing the manuscript.

\subsection*{Additional information}

Supplementary information is available in the online version of the paper. Reprints and
permission information is available online at http://www.nature.com/reprints. Correspondence
and requests for materials should be addressed to R.A. and G.M.

\subsection*{Competing financial interests}

The authors declare no competing financial interests.


\begin{thebibliography}{10}

\bibitem{Hell92a}Hell, S. \& Stelzer, E.H.K. Fundamental improvement of resolution with a 4Pi-confocal fluorescence microscope using two-photon excitation. \emph{Opt. Commun.} {\bf 93,} 277--282 (1992).

\bibitem{Hell92b}Hell, S. \& Stelzer, E.H.K. Properties of a 4Pi confocal fluorescence microscope. \emph{J. Opt. Soc. Am. A} {\bf 9,} 2159--2166 (1992).

\bibitem{Gu94}Gu, M. \& Sheppard, C.J.R. Three-dimensional transfer functions in 4Pi confocal microscopes. \emph{J. Opt. Soc. Am. A} {\bf 11,} 1619--1627 (1994).

\bibitem{Lanni86}Lanni, F. [Standing-wave fluorescence microscopy] \emph{Applications of Fluorescence in the Biomedical Sciences} [Taylor, D.L., Waggoner, A.S., Lanni, F., Murphy, R.F. \& Birge, R.R. (eds.)] [505--521] (Alan R. Liss, Inc., New York, 1986).


\bibitem{Lanni86p}Lanni, F., Taylor, D.L. \& Waggoner, A.S. US Patent No. 4,621,911 (1986).

\bibitem{Bailey93}Bailey, B., Farkas, D.L., Taylor, D.L. \& Lanni, F. Enhancement of axial resolution in fluorescence microscopy by standing-wave excitation. \emph{Nature} {\bf 366,} 44--48 (1993).

\bibitem{Elsayad2013}Elsayad, K. \emph{et al.} Spectrally coded optical nanosectioning (SpecON) with biocompatible metal--dielectric-coated substrates. \emph{Proc. Natl Acad. Sci. USA} {\bf 110,} 20069--20074 (2013).

\bibitem{Amos2012}Amos, W.B., McConnell, G. \& Wilson, T. Confocal microscopy. \emph{Comprehensive Biophysics}, ed Egelman, E. (Elsevier, Amsterdam, 2012).



\bibitem{Egner2005}Egner, A. \& Hell, S.W. Fluorescence microscopy with super-resolved sections. \emph{Trends Cell Biol.} {\bf 15,} 207--215 (2005).

\bibitem{Freimann97}Freimann, R., Pentz, S. \& H\"{o}rler, H. Development of a standing-wave fluorescence microscope with high nodal plane flatness. \emph{J. Microsc.} {\bf 187,} 193--200 (1997).

\bibitem{Wiener1890}Wiener, O. Stehende lichtwellen und die schwingungsrichtung polarisirten lichtes. \emph{Ann. Phys.} {\bf 276,} 203--243 (1890).

\bibitem{Ditchburn58}Ditchburn, R.W. \emph{Light} (Blackie \& Son Ltd., London, 1958).

\bibitem{Jenkins76}Jenkins, F.A. \& White, H.E. \emph{Fundamentals of Optics}, 4th ed (McGraw-Hill Book Company, Inc., New York, 1976).

\bibitem{Hecht98}Hecht, E. \emph{Optics}, 3rd ed (Addison Wesley Longman, Inc., Reading, Massachusetts, 1998).



\bibitem{Leica5nm2014}Leica TCS SP5 II Technical Documentation (2009). Available at www.leica-microsystems.com. Date of access: 16/09/2014.



\bibitem{Rogers77}Rogers, G.L. A geometrical approach to moir\'{e} pattern calculations. \emph{Optica Acta} {\bf 24,} 1--13 (1977).

\bibitem{Myers2004}Barrett, HH. \& Myers, K.J. \emph{Foundations of Image Science} (John Wiley \& Sons, Inc., Hoboken, New Jersey, 2004).

\bibitem{Stupp2008}Brennesholtz, M.S. \& Stupp, E.H. \emph{Projection Displays}, 2nd ed (John Wiley \& Sons Ltd., West Sussex, 2008).



\bibitem{Selenyi11}Selenyi, P. \"{U}ber Lichtzerstreuung im Raume Wienerscher Interferenzen und neue, diesen reziproke Interferenzerscheinungen. \emph{Ann. Phys.} {\bf 340,} 444--460 (1911).

\bibitem{Selenyi38}Selenyi, P. Herstellung und eigenschaften weitwinkliger optischer interferenzerscheinungen. \emph{Zeits. f. Physik.} {\bf 108,} 401--407 (1938).

\bibitem{Selenyi39}Selenyi, P. Wide-angle interferences and the nature of the elementary light sources. \emph{Phys. Rev.} {\bf 56,} 477--479 (1939).


\bibitem{Drexhage68}Drexhage, K.H., Kuhn, H. \& Schaefer, F.P. Variation of the fluorescence decay time of a molecule in front of a mirror. \emph{Ber. Bunsenges. Phys. Chem.} {\bf 72,} 329 (1968).

\bibitem{Drexhage70a}Drexhage, K.H. Influence of a dielectric interface on fluorescence decay time. \emph{J. Lumin.} {\bf 1--2,} 693--701 (1970).

\bibitem{Drexhage70b}Drexhage, K.H. Monomolecular layers and light. \emph{Sci. Am.} {\bf 222,} 108--119 (1970).

\bibitem{Drexhage74}Drexhage, K.H. [Interaction of light with monomolecular dye layers] \emph{Progress in Optics XII} [Wolf, E. (ed.)] [165--231] (North-Holland, Amsterdam, 1974).




\bibitem{Lukosz77i}Lukosz, W. \& Kunz, R.E. Light emission by magnetic and electric dipoles close to a plane interface. I. Total radiated power. \emph{J. Opt. Soc. Am.} {\bf 67,} 1607--1615 (1977).

\bibitem{Lukosz77ii}Lukosz, W. \& Kunz, R.E. Light emission by magnetic and electric dipoles close to a plane dielectric interface. II. Radiation patterns of perpendicular oriented dipoles. \emph{J. Opt. Soc. Am.} {\bf 67,} 1615--1619 (1977).

\bibitem{Lukosz79}Lukosz, W. Light emission by magnetic and electric dipoles close to a plane dielectric interface. III. Radiation patterns of dipoles with arbitrary orientation. \emph{J. Opt. Soc. Am.} {\bf 69,} 1495--1503 (1979).

\bibitem{Axelrod87}Hellen, E.H. \& Axelrod, D. Fluorescence emission at dielectric and metal-film interfaces. \emph{J. Opt. Soc. Am. B} {\bf 4,} 337--350 (1987).

\bibitem{Axelrod89}Hellen, E.H. \& Axelrod, D. [Emission of fluorescence at an interface] \emph{Methods in Cell Biology, Vol. 30: Fluorescence Microscopy of Living Cells in Culture, Part B: Quantitative Fluorescence Microscopy - Imaging and Spectroscopy} [Taylor, D.L. \& Wang, Y.-L. (eds.)] [399--416] (Academic Press, Inc., San Diego, California, 1989).



\bibitem{Lambacher96}Lambacher, A. \& Fromherz, P. Fluorescence interference-contrast microscopy on oxidized silicon using a monomolecular dye layer. \emph{Appl. Phys. A} {\bf 63,} 207--216 (1996).



\bibitem{Dow2014}Density of Glycerine-Water Solutions (2013). Available at www.dow.com. Date of access: 25/02/2014.

\bibitem{Comar2014}Optical Materials Technical Information (2013). Available at www.comaroptics.com. Date of access: 25/04/2014.


\bibitem{Schlessinger77}Schlessinger, J., Axelrod, D., Koppel, D.E., Webb, W.W. \& Elson, E.L. Lateral transport of a lipid probe and labeled proteins on a cell membrane. \emph{Science} {\bf 195,} 307--309 (1977).

\bibitem{Chang89}Chang, D.C. Cell poration and cell fusion using an oscillating electric field. \emph{Biophys. J.} {\bf 56,} 641--652 (1989).

\bibitem{Chen2008}Chen, H. \emph{et al.} Release of hydrophobic molecules from polymer micelles into cell membranes revealed by F\"{o}rster resonance energy transfer imaging. \emph{Proc. Natl Acad. Sci. USA} {\bf 105,} 6596--6601 (2008).

\bibitem{Hod2010}Hod, E.A. \emph{et al.} Transfusion of red blood cells after prolonged storage produces harmful effects that are mediated by iron and inflammation. \emph{Blood} {\bf 115,} 4284--4292 (2010).

\bibitem{Agmon95}Agmon, A., Yang, L.T., Jones, E.G. \& O'Dowd, D.K. Topological precision in the thalamic projection to neonatal mouse barrel cortex. \emph{J. Neurosci.} {\bf 15,} 549--561 (1995).

\bibitem{Webb83}Bloom, J.A. \& Webb, W.W. Lipid diffusibility in the intact erythrocyte membrane. \emph{Biophys. J.} {\bf 42,} 295--305 (1983).





\bibitem{Greer2009}Greer, J.P. \emph{et al.} \emph{Wintrobe's Clinical Hematology, Vol 1}, 12th ed (Lippincott Williams and Wilkins, Philadelphia, 2009).

\bibitem{Harris2001}Harris, F.M., Smith, S.K. \& Bell, J.D. Physical properties of erythrocyte ghosts that determine susceptibility to secretory phospholipase A$_{2}^{*}$. \emph{J. Biol. Chem.} {\bf 276,} 22722--22731 (2001).

\bibitem{Barer54}Barer, R. \& Tkaczyk, S. Refractive index of concentrated protein solutions. \emph{Nature} {\bf 173,} 821--822 (1954).


\bibitem{Tse99}Tse, W.T. \& Lux, S.E. Red blood cell membrane disorders. \emph{Br. J. Haematol.} {\bf 104,} 2--13 (1999).

\bibitem{LimHW2002}Lim, H.W.G., Wortis, M. \& Mukhopadhyay, R. Stomatocyte--discocyte--echinocyte sequence of the human red blood cell: Evidence for the bilayer--couple hypothesis from membrane mechanics. \emph{Proc. Natl Acad. Sci. USA} {\bf 99,} 16766--16769 (2002).

\bibitem{Gallagher2005}Gallagher, P.G. Red cell membrane disorders. \emph{Hematology Am. Soc. Hematol. Educ. Program} {\bf 1,} 13--18 (2005).

\bibitem{Dvorak75}Dvorak, J.A., Miller, L.H., Whitehouse, W.C. \& Shiroishi, T. Invasion of erythrocytes by malaria merozoites. \emph{Science} {\bf 187,} 748--750 (1975).

\bibitem{Pinder98}Pinder, J.C. \emph{et al.} Actomyosin motor in the merozoite of the malaria parasite, \emph{Plasmodium falciparum}: implications for red cell invasion. \emph{J. Cell Sci.} {\bf 111,} 1831--1839 (1998).

\bibitem{Salmon2001}Salmon, B.L., Oksman, A. \& Goldberg, D.E. Malaria parasite exit from the host erythrocyte: A two-step process requiring extraerythrocytic proteolysis. \emph{Proc. Natl Acad. Sci. USA} {\bf 98,} 271--276 (2001).

\bibitem{Stubbs2005}Stubbs, J. \emph{et al.} Molecular mechanism for switching of \emph{P. falciparum} invasion pathways into human erythrocytes. \emph{Science} {\bf 309,} 1384--1387 (2005).

\bibitem{Gilson2009}Gilson, P.R. \& Crabb, B.S. Morphology and kinetics of the three distinct phases of red blood cell invasion by \emph{Plasmodium falciparum} merozoites. \emph{Int. J. Parasitol.} {\bf 39,} 91--96 (2009).

\bibitem{Riglar2011}Riglar, D.T. \emph{et al.} Super-resolution dissection of coordinated events during malaria parasite invasion of the human erythrocyte. \emph{Cell Host Microbe} {\bf 9,} 9--20 (2011).

\end{thebibliography}

\newpage
\section*{Figures}

\end{document}